\documentclass[10pt, onecolumn, conference]{IEEEtran}

\usepackage[T1]{fontenc}
\usepackage[utf8]{inputenc}
\usepackage[english]{babel}
\usepackage{array}
\usepackage{graphicx}
\usepackage{booktabs}
\usepackage{amsmath}
\usepackage{amssymb}
\usepackage{multirow}
\usepackage{xcolor}
\usepackage{url}
\usepackage{float}  
\usepackage{hyperref}

\providecommand{\doi}[1]{\url{https://doi.org/#1}}

\newcommand{\anontool}{AnonShield}

\begin{document}

\title{Decomposing Memorization Reduction in Privacy-Preserving Fine-Tuning of SLMs for CSIRTs}

\author{
\IEEEauthorblockN{Cristhian Kapelinski\IEEEauthorrefmark{1}, Diego Kreutz\IEEEauthorrefmark{1}}
\IEEEauthorblockA{\IEEEauthorrefmark{1}Federal University of Pampa (UNIPAMPA), Alegrete, Brazil}
\IEEEauthorblockA{\{cristhianavilla.aluno, diegokreutz\}@unipampa.edu.br}
}

\maketitle

\begin{abstract}
CSIRTs increasingly fine-tune language models on vulnerability scans, yet
such records expose internal topology and create privacy risk under
GDPR/LGPD. We present the first empirical study of how DP-SGD and HMAC
pseudonymization compose when fine-tuning $1$--$3$B SLMs on structured
CSIRT data. Across 96 LoRA adapters over four SLMs, four regimes (raw,
QLoRA with large-lot batching, DP-SGD at
$\varepsilon\!\in\!\{2,8\}$), and 20 canaries audited via four extraction
attacks plus a dual attack on the HMAC slug:
\textbf{(i)}~matched-update controls reproduce the memorization
reduction in privacy-preserving fine-tuning by reducing optimizer
updates alone ($66$--$132\%$ of the gap, mean $100\%$, n${=}3$ seeds,
4 models); DP-SGD adds $(\varepsilon,\delta)$ without further reduction.
\textbf{(ii)}~HMAC pseudonymization removes the original identifier from
the exposure surface ($40$--$61\%$) without creating a secondary target:
slug exposure remains within $\pm0.65$ bits of the chance floor.
\textbf{(iii)}~F1 stays in $[0.19,0.28]$ across all 96 adapters and
four-shot prompting, indicating a budget-conditional gap: under the
evaluated regime, $1$--$3$B SLMs do not reach operational F1.\footnote{\url{https://anonymous.4open.science/r/csirt-slm-memorization-5BBC}}
\end{abstract}

\begin{IEEEkeywords}
Memorization, Differential Privacy, Pseudonymization, Small Language Models, CSIRT.
\end{IEEEkeywords}

\section{Introduction}
\label{sec:intro}

The operational scale of incident response has outpaced the public
infrastructure that historically supported triage. With CVE submissions
up $263\%$ between 2020 and 2025, on 15~April~2026 NIST adopted a tiered
policy enriching only high-priority CVEs (CISA KEV, federal systems,
EO~14028 critical software) on a one-business-day target, while the long
tail is marked \emph{Not Scheduled} and may never receive severity
scores, CWE, or CPE
identifiers~\cite{NIST2026Update}, shifting that burden onto the local
CSIRT and motivating automation beyond rule-based pipelines.

CSIRTs (Computer Security Incident Response Teams) increasingly turn to language models for triage, classification,
and remediation
guidance~\cite{AnonRef2,Habibzadeh2025}. Fine-tuning
on operational records embeds identifiers (IPs, hostnames, asset names,
certificate fingerprints) that map an organization's internal topology,
in tension with GDPR and LGPD. A decade of work establishes that planted
canaries are extractable~\cite{Carlini2019,Lukas2023,Hayes2025},
that memorization scales with capacity, duplication, and prompt
length~\cite{Carlini2023,Kandpal2022}, and that fine-tuning, including
PEFT, retains measurable
leakage~\cite{Mireshghallah2022,WangLi2025}; concurrent work
cautions that some apparent extraction tracks generalization rather than
memorization~\cite{Duan2024MIA,Wang2025Generalization}, a distinction we honor by measuring
exposure against in-distribution variants.

Three families of mitigations dominate the literature, none sufficient in
isolation~\cite{Tramer2022,Lukas2023}: DP-SGD~\cite{Abadi2016,Yu2022,Li2022}
delivers formal $(\varepsilon,\delta)$ guarantees but couples
non-trivially with parameter-efficient fine-tuning
(PEFT)~\cite{Hu2022} and degrades utility at strict budgets;
pseudonymization removes identifiers before training, yet naive scrubbing
leaves residual leakage~\cite{Lukas2023}; output filters
fall under prompt-level perturbations and style
transfer~\cite{Ippolito2023}. The three act at distinct points
(pseudonymization at the data layer, DP-SGD at the optimizer, filtering
at decoding), neither subsume nor compose trivially, and how they
interact in fine-tuning is unresolved.

\noindent\textit{Three gaps converge on the $1$--$3$B regime.} First,
prior DP-SGD evaluations on language
models~\cite{Yu2022,Li2022,Anil2022} report predominantly on
encoder backbones (BERT, RoBERTa) or $\leq$1.5B GPT-2 decoders, on
general-purpose benchmarks (GLUE, E2E, Web-EN), not on operational
domain-specific records. Second, work that measures memorization on
language
models~\cite{Carlini2019,Carlini2021,Mireshghallah2022,Lukas2023,Jagielski2023}
either does not evaluate formal defenses or evaluates one at a time,
leaving open how DP-SGD composes with cryptographic pseudonymization on
the same canary set. Third, recent work probing LMs in CSIRT/SOC
settings~\cite{AnonRef1,AnonRef2,Rigaki2024Hackphyr,Krishna2024AttackQA,Hays2025SOC,Kramer2025SOUPS}
either sits at $\geq$7B with prompt engineering on free-text narratives
or fine-tunes locally without formal privacy mechanisms or memorization
measurement. The $1$--$3$B fine-tuning regime over structured CSIRT
records under composed defenses is therefore the intersection no prior
work occupies, and it matches the on-premise hardware constraints of
most operational CSIRTs.

\noindent\textit{Contributions.}
We present the first empirical study of memorization and classification
utility in fine-tuning $1$--$3$B SLMs over composed protection regimes on
CSIRT vulnerability records:
\textbf{(C1)} a four-regime ladder (raw, QLoRA + large-lot batching,
DP-SGD at $\varepsilon\!\in\!\{2,8\}$) over four SLMs (Gemma~3, Qwen3,
Llama~3.2, VaultGemma), yielding 96 LoRA adapters under cross-seed audit;
matched-update controls reproduce $66$--$132\%$ of the raw-to-stack
reduction (mean $100\%$, n${=}3$) by reducing optimizer updates alone, with
DP-SGD adding $(\varepsilon,\delta)$ without further reduction (\S\ref{sec:exp1});
\textbf{(C2)} a dual extraction attack on HMAC pseudonymization
(\anontool~\cite{AnonShield2026}) querying original and slug, showing
removal of the original from the exposure surface ($40$--$61\%$ in the raw regime)
without creating a secondary target ($\pm0.65$ bits from the chance
floor across 16 model$\times$variant cells; \S\ref{sec:exp2});
\textbf{(C3)} a budget-conditional viability gap from logit fine-tuning and
four-shot prompting: under the evaluated training regime, $1$--$3$B SLMs
stay below operational F1 (F1$\in[0.19,0.28]$ across 96 adapters; \S\ref{sec:utility}). We also
highlight a deployment hazard for DP-SGD on multimodal-by-default
architectures (\S\ref{sec:models}) and release the dataset, pipeline, and
raw log-probabilities for re-analysis under probabilistic extraction
metrics~\cite{Hayes2025}.\footnotemark[1]

\section{Related Work}
\label{sec:related_work}

Table~\ref{tab:related_work} positions our work along six descriptive
axes; we develop the comparison in three families: memorization
measurement, privacy-preserving training, and SLM/LLM use in CSIRT
operations. 

\begin{table}[H]
\centering
\caption{Related work along six descriptive axes. \emph{Approach}: model
adaptation/training mode. \emph{Defense}: privacy mechanism, with formal
guarantee in parentheses ($(\varepsilon,\delta)$=DP, PRF=pseudo-random
function). \emph{Memo.\ probe}: memorization protocol. ``---''=N/A.}
\label{tab:related_work}
\renewcommand{\arraystretch}{1.05}
\resizebox{\textwidth}{!}{%
\scriptsize
\begin{tabular}{|>{\raggedright\arraybackslash}p{2.6cm}|>{\raggedright\arraybackslash}p{1.9cm}|>{\raggedright\arraybackslash}p{2.4cm}|>{\raggedright\arraybackslash}p{2.7cm}|>{\raggedright\arraybackslash}p{2.4cm}|>{\raggedright\arraybackslash}p{2.0cm}|}
\hline
\textbf{Work} & \textbf{Approach} & \textbf{Models} & \textbf{Defense (guarantee)} & \textbf{Memo.\ probe} & \textbf{Domain} \\
\hline
\multicolumn{6}{l}{\emph{\textbf{Memorization measurement}}} \\
\hline
Carlini et al.~\cite{Carlini2019} & Pretraining &LSTM (Smart Compose) & --- & Exposure (Secret Sharer) & Email \\
\hline
Carlini et al.~\cite{Carlini2021} & Pretraining &GPT-2 (124M--1.5B) & --- & Verbatim extraction & Web-EN \\
\hline
Mireshghallah et al.~\cite{Mireshghallah2022} & Fine-tuning &GPT-2 ($\sim$124M) & --- & MIA recall + Exposure & Wiki / PTB / Enron \\
\hline
Lukas et al.~\cite{Lukas2023} & Fine-tuning &GPT-2 (Small--XL) & scrub + DP-SGD ((\(\varepsilon,\delta\))) & PII extraction / reconstruction / inference & ECHR / Enron / Yelp \\
\hline
Jagielski et al.~\cite{Jagielski2023} & Pre/Fine-tuning &110M LM + ResNet & --- & Forgetting over training & C4/vision \\
\hline
Kandpal et al.~\cite{Kandpal2024UserInf} & Fine-tuning &LLM (GPT-Neo) & --- & User-level inference & Reddit \\
\hline
\multicolumn{6}{l}{\emph{\textbf{Privacy-preserving training}}} \\
\hline
Anil et al.~\cite{Anil2022} & Pretraining &BERT-Large (340M) & DP-SGD ((\(\varepsilon,\delta\))) & --- & Wiki + Books \\
\hline
Yu et al.~\cite{Yu2022} & Fine-tuning &RoBERTa, GPT-2 & DP-SGD + LoRA/Adapter ((\(\varepsilon,\delta\))) & --- & GLUE / E2E \\
\hline
Li et al.~\cite{Li2022} & Fine-tuning &RoBERTa, GPT-2 & DP-SGD + Ghost Clipping ((\(\varepsilon,\delta\))) & --- & GLUE / E2E \\
\hline
Steinke et al.~\cite{Steinke2023} & Training &WideResNet & DP-SGD audit (1-run) & DP audit & CIFAR-10 \\
\hline
Sinha et al.~\cite{Sinha2025} & Pretraining &VaultGemma 1B & DP-PT ((\(\varepsilon,\delta\))) & --- & Web-EN \\
\hline
Nitz et al.~\cite{Nitz2025SAPPAN} & Federated &DGA/malware classifiers & FL + crypto (SAPPAN) & --- & CTI (EU) \\
\hline
\multicolumn{6}{l}{\emph{\textbf{LLMs/SLMs in SOC/CSIRT operations}}} \\
\hline
Rigaki et al.~\cite{Rigaki2024Hackphyr} & Fine-tuning &Zephyr-7B (local) & local deployment (privacy-motivated) & --- & Network security red-team \\
\hline
Badrinath Krishna~\cite{Krishna2024AttackQA} & Fine-tuning + RAG &Llama 3 8B / 70B & none (public MITRE KB) & --- & SOC Q\&A (MITRE ATT\&CK) \\
\hline
Singh et al.~\cite{Hays2025SOC} & Prompt eng.\ & GPT-4-0613 & none (cloud API study) & --- & SOC narratives (field) \\
\hline
Kramer et al.~\cite{Kramer2025SOUPS} & Prompt eng.\ & Gemini 1.5 Flash & none (cloud API study) & --- & IR narratives (qual.) \\
\hline
Severo et al.~\cite{AnonRef1} & Prompt eng.\ & Gemini 2.0 Flash, GPT-4, Grok, Llama 3.1 70B & anonymization at input & --- & CSIRT narratives (BR) \\
\hline
\multicolumn{6}{l}{\emph{\textbf{}}} \\
\hline
\textbf{This work} & \textbf{Fine-tuning} & \textbf{1--3B SLMs (4 models) + QLoRA} & \textbf{DP-SGD + HMAC ((\(\varepsilon,\delta\))+PRF)} & \textbf{Exp.\ + 3\(\times\)AUC + Min-K\%++} & \textbf{CSIRT vuln.\ scan (BR)} \\
\hline
\end{tabular}}
\end{table}

\noindent\textit{Memorization measurement.}
The dominant probe in Tab.~\ref{tab:related_work} is canary
exposure~\cite{Carlini2019} and its variants: verbatim
extraction~\cite{Carlini2021}, threshold detectors
(Min-K\%~\cite{Shi2024}, Min-K\%++~\cite{Zhang2025}), probabilistic
extraction~\cite{Hayes2025}, and MIA-style
recall~\cite{Lukas2023}. Two
patterns emerge. First, all works evaluate a single mechanism in
isolation (scrub or DP-SGD~\cite{Lukas2023}, fine-tuning
regime~\cite{Mireshghallah2022}, duplication~\cite{Jagielski2023},
user-level boundaries~\cite{Kandpal2024UserInf}), never two defenses on
the same canary set. Second, domains are general-English corpora (Web,
Wiki, Reddit, Enron); none studies structured records with
schema-fixed identifiers. Duan et al.~\cite{Duan2024MIA} show that
loss-based MIA can confound memorization with distribution shift, which
we avoid via in-distribution variant pools (\S\ref{sec:canaries}).
Mireshghallah et al.~\cite{Mireshghallah2022} extract canaries from
fine-tuned LLMs, and Zanella-B\'eguelin et al.~\cite{ZanellaBeguelin2020}
study leakage from model updates analogous to LoRA adapters.

\noindent\textit{Privacy-preserving training.}
The \emph{Defense} column of Tab.~\ref{tab:related_work} spans two axes.
DP-SGD~\cite{Abadi2016} provides formal $(\varepsilon,\delta)$ with
utility cost, mitigated via adapters~\cite{Yu2022,Li2022}, scale
extensions~\cite{Anil2022,Raisa2024Subsampling}, tighter audits~\cite{Nasr2021,Steinke2023},
and DP pretraining~\cite{Sinha2025}, whose guarantees cover pretraining
data only~\cite{Tramer2022}. The data-side axis substitutes or scrubs
identifiers~\cite{Lukas2023}; decoding filters form a
weaker axis~\cite{Ippolito2023}. These axes are rarely composed:
Lukas et al.~\cite{Lukas2023} evaluates both, but only on English PII.
How a cryptographic PRF defense composes with DP-SGD on a shared canary
set remains open.

\noindent\textit{LLM/SLM use in CSIRT operations.}
Two deployment families dominate. Cloud LLMs at $\geq$7B with prompt
engineering on free text~\cite{Hays2025SOC,Kramer2025SOUPS,AnonRef1,Lin2025}
trade utility for provider-side exposure; local fine-tuning at
$7$--$70$B~\cite{Rigaki2024Hackphyr,Krishna2024AttackQA} addresses
sovereignty but omits privacy and memorization analysis. Surveys cover
applications from triage to MITRE ATT\&CK
classification~\cite{Habibzadeh2025}, and
SAPPAN~\cite{Nitz2025SAPPAN} studies federated CTI sharing.
\anontool~\cite{AnonShield2026} provides schema-driven HMAC
pseudonymization; the $1$--$3$B on-premise regime remains unexplored.

\noindent\textit{Gap.}
While combining pseudonymization and differential privacy is known,
their empirical composition in SLM fine-tuning on structured CSIRT data
is unstudied. No prior work compares DP-SGD with deterministic
pseudonymization on the same canary set in the $1$--$3$B regime, nor
measures memorization of pseudonyms. We close this gap via a dual-attack
protocol evaluating both original identifiers and HMAC slugs across
96 LoRA adapters.

\section{Threat Model and Methodology}
\label{sec:method}

\subsection{Threat Model}
\label{sec:threat}

We consider a black-box adversary with query access to a fine-tuned LoRA
adapter, able to submit arbitrary prefixes and observe token-level
log-probabilities (equivalently, per-token losses) over a fixed
vocabulary. The adversary has no access to the HMAC key, training seeds,
gradients, optimizer state, or intermediate activations, and cannot
observe other constituents' queries. This matches the intended
deployment: CSIRTs distribute adapters to constituents for local triage
via standard inference runtimes
(vLLM~\cite{Kwon2023vLLM}, \texttt{llama.cpp}, HuggingFace Transformers),
all of which expose the log-probability channel through their standard
APIs. White-box gradient inversion~\cite{Geiping2020}, training-time
membership inference~\cite{Shokri2017}, and model
extraction~\cite{Carlini2024Stealing} require capabilities outside this
surface and are out of scope.

\subsection{Models}
\label{sec:models}

We evaluate four open-weight SLMs in the $1$--$3$B range from three
independent families, controlling for tokenizer and pretraining-corpus
idiosyncrasies in the memorization signal: Gemma~3
1B-IT~\cite{Gemma3}, Qwen3 1.7B~\cite{Qwen3}, Llama~3.2 3B-Instruct~\cite{Llama3}, and
VaultGemma~1B~\cite{Sinha2025}. VaultGemma, the largest open-weight LM
trained from inception with a formal DP guarantee, adds an orthogonal axis: it tests how DP-at-pretraining composes with
fine-tuning-time defenses against canaries injected \emph{after}
pretraining, a regime in which the scope principle of DP predicts no
protection from the pretraining-time guarantee~\cite{Tramer2022}. The
$1$--$3$B ceiling is bounded by hardware (RTX~5060~Ti 16~GB, RTX~3060
12~GB): under QLoRA + DP-SGD with Ghost Clipping~\cite{Li2022}, 3B
saturates 16~GB VRAM at sequence length 768. We treat this as a
worst-case feasibility test for on-premise CSIRT deployments;
$\geq$7B is future work.

\noindent\textit{Multimodal SLMs are excluded by a DP-SGD compatibility
constraint, not by design.}
Recent multimodal SLMs (e.g., Gemma~4 E2B, Qwen3-VL) carry vision and
audio towers whose attention modules share \texttt{q\_proj},
\texttt{k\_proj}, \texttt{v\_proj}, \texttt{o\_proj} names with the
text path. LoRA target matching attaches adapters to all four
projections in every tower; under text-only inputs, non-text towers
receive no gradient and Opacus'~\cite{Yousefpour2021Opacus}
\texttt{grad\_sample} stays un-populated, aborting the optimizer step
(\emph{``Per sample gradient is not initialized''}). The standard
workaround is per-architecture exclusion regexes; we sidestep it by
restricting the scan to text-only checkpoints and flag the interaction
as a deployment hazard for future DP-SGD evaluation on multimodal
foundation models. Pseudonymization is unaffected: it is a data-side
preprocessing pass, independent of the downstream architecture.

\subsection{Datasets}
\label{sec:datasets}

The reference corpus is $70{,}951$ vulnerability records from a Tenable
scanner at CSIRT-RNP, the Brazilian NREN incident response team. We focus
on \emph{vulnerability scan records} (asset metadata, CVE/CVSS, plugin
family, technical descriptions), distinct from the analyst-authored
narratives more commonly studied in CSIRT
NLP~\cite{AnonRef1}. Records carry FQDNs, IPv4/IPv6 and MAC
addresses, asset names, and OS strings mapping the internal topology of
academic and federal networks; redistribution would violate LGPD and
constituent confidentiality.

\noindent\textit{Synthetic surrogate.}
\emph{Mock~CAIS} replicates $70{,}951$ Tenable-schema records in three
deterministic stages. \emph{(i)} streams the real corpus for non-PII
histograms, categorical proportions, per-record CVE/CWE/CPE/tag counts,
and the $9{,}556$ real CVE-IDs to exclude. \emph{(ii)} draws one CVE per
record from a public NVD mirror over non-excluded IDs, samples
categoricals empirically, and draws identifiers from
RFC~5737/3849/1918/4193 ranges and Faker pt\_BR, guaranteeing zero
collision with routable infrastructure. \emph{(iii)} validates fidelity
following Snoke et al.~\cite{Snoke2018} via marginal tests (KS,
Anderson-Darling, $\chi^2$ with BH-FDR~\cite{Benjamini1995}), joint tests
(MMD-RBF, C2ST, pMSE), the manifold metrics of Alaa et
al.~\cite{Alaa2022}, and lexical diversity over $70{,}951$ paired
records. The pipeline is fully seeded. All six categorical fields pass
$\chi^2$ with $p>0.5$; $\mathrm{MMD}^2_{\text{RBF}}=0.146$; manifold
authenticity $0.93$; CVE-ID overlap is zero by construction. C2ST
AUC$=0.98$ is driven by design-level gaps in
\texttt{description\_len}/\texttt{solution\_len} (Tenable prose vs.\ NVD
raw text), neutral here since canary exposure is measured against
in-distribution variants.

\noindent\textit{Splits and pseudonymization.}
Three deterministic subsets: $3{,}000$ stratified organic training
records (750 per CVSS bucket), $300$-record canary held-out, $600$-record
utility held-out (Low~150 / Med~200 / High~150 / Critical~100;
\S\ref{sec:utility}); $200$ planted canaries yield the $3{,}200$-record
training corpus. For Experiment~2, records traverse
\anontool~\cite{AnonShield2026}, replacing each structured identifier
with \texttt{[TYPE\_<8 hex>]}, the leading bytes of
$\mathrm{HMAC\text{-}SHA256}_K(\text{TYPE}\mathbin{\Vert}\text{value})$;
substitution is deterministic under a fixed key, reversible only with
the key, and covers all FQDNs, IPv4/IPv6 and MAC addresses, asset names,
and OS strings. Canaries are planted \emph{before} pseudonymization, so
the model trains on the slug while the original is retained as ground
truth for the dual attack (\S\ref{sec:exp2}). Residual identifiers in
free-text \texttt{description}/\texttt{solution} are out of scope.

\subsection{Canaries and Attacks}
\label{sec:canaries}

We plant 20 canaries across five identifier types (IPv4, IPv6, FQDN,
MAC, asset), split into \emph{Class~A (anomalous)} from RFC~5737/3849
ranges and \emph{Class~B (realistic)} from RFC~1918/4193 and Brazilian
hostname patterns to test anomaly- vs.\ realism-driven leakage. Each
canary is replicated $K{=}10$ times~\cite{Carlini2019}, yielding 200
planted records over $3{,}000$ organic ones; per canary, a pool of 100
syntactically valid \emph{variants} (same /24, suffix, or OUI) forces the
metric to capture memorization of the \emph{specific} value. In
Experiment~2, canaries are injected in plain text and traverse
\anontool, so the model sees the slug while the original is retained for
the dual attack (\S\ref{sec:exp2}).

\noindent\textit{Exposure metric.}
Leakage of canary $c$ relative to its 100-variant pool $V$ is
\begin{equation}
\mathrm{Exp}(c) = \log_{2}(|V|+1) - \log_{2}\mathrm{rank}(c),
\label{eq:exposure}
\end{equation}
with $\mathrm{rank}(c)$ the position of $c$'s loss (rank~1 = most
memorized)~\cite{Carlini2019}. With $|V|{=}100$,
$\mathrm{Exp}\in[0,\log_{2}101]$ and an unseen $c$ yields
$\log_{2}(101/50)\approx1$ bit (floor). We retain the rank-based form
for comparability and release raw log-probabilities for re-analysis
under probabilistic relaxations~\cite{Hayes2025}.

\noindent\textit{Attack suite.}
Per adapter we run four attacks:
(i)~\textbf{Carlini exposure}~\cite{Carlini2019};
(ii)~\textbf{Loss-MIA AUC}~\cite{Yeom2018};
(iii)~\textbf{Loss-canary AUC} (this work), a token-restricted variant
of loss-based MIA~\cite{Yeom2018,Lukas2023};
(iv)~\textbf{Min-K\%++}~\cite{Zhang2025} with $k{=}20\%$. They differ in
loss aggregation and provide concordant evidence. Loss-based MIA may
confound memorization with distribution shift~\cite{Duan2024MIA}; our
variant pool is in-distribution by construction. Statistical analysis
uses Wilcoxon signed-rank (paired by canary, seed), Friedman tests,
Cliff's $\delta$, and BH-FDR~\cite{Benjamini1995} at $q{=}0.05$.
$\mathrm{Exp}(c)$ is reported as per-canary means
(Tab.~\ref{tab:exp1_main}) or per-class means (Tab.~\ref{tab:exp2_main}).

\subsection{Fine-Tuning Regimes}
\label{sec:variants}

We evaluate four regimes as a controlled ladder, each adding one mechanism
to the previous to isolate its marginal contribution. All use
LoRA~\cite{Hu2022} with $r{=}16$, $\alpha{=}32$, dropout~$0.05$, target
modules $\{q,k,v,o\}_{\text{proj}}$. \textbf{V0 (raw)}: bf16 weights with
HuggingFace \texttt{SFTTrainer}, unprotected baseline.
\textbf{V1 (stack)}: V0 plus QLoRA NF4 4-bit quantization~\cite{Dettmers2023}
and uniform-shuffle large-lot batching (effective lot $32$ via gradient
accumulation); no privacy guarantee, but matches the stack typically
co-deployed with DP-SGD, separating its empirical effect from the formal
$(\varepsilon,\delta)$ one.
\textbf{V2 ($\varepsilon{=}8$)}: V1 plus Opacus
PrivacyEngine~\cite{Yousefpour2021Opacus} with Ghost
Clipping~\cite{Li2022}. \textbf{V3 ($\varepsilon{=}2$)}: V2 under a
conservative budget. Both DP regimes use $\delta{=}10^{-5}\ll 1/n$ for
$n{=}3200$, clipping norm $C{=}1.0$, sampling rate $q{=}0.01$, and the PRV
accountant~\cite{Gopi2021PRV}; $\sigma$ is tuned per $\varepsilon$ over 300
steps ($0.547$ at $\varepsilon{=}8$; $0.885$ at $\varepsilon{=}2$). All
regimes train 3 epochs with AdamW, lr $10^{-4}$ cosine-annealed, and max
sequence length $1024$ (reduced to $768$ for Llama~3B in V2/V3 to fit
Ghost Clipping; canary identifiers ($\leq30$ chars) fit regardless, only
\texttt{description}/\texttt{solution} tails are truncated). Each regime
runs on each model for 3 seeds, yielding 48 adapters per experiment (96
total) at $\approx 222$~GPU-hours on the setup of
Section~\ref{sec:models}. Two intermediate ablation variants
$A$ (lot~32, bf16, uniform shuffle) and $B$ (lot~4, NF4, uniform
shuffle) further isolate the marginal contributions of large-lot
batching and 4-bit quantization within the V0$\to$V1 stack
(\S\ref{sec:exp1}); they are run on all four models, cross-seed (n${=}3$).

\section{Memorization Across Protection Regimes}
\label{sec:exp1}

The first experiment measures memorization of canaries in raw
(non-anonymized) records across V0--V3, training 3 epochs over $3{,}200$
records. Empirical $\varepsilon$ from the PRV accountant~\cite{Gopi2021PRV}
matches targets within $\pm 0.01$ across 24 DP runs. Audit checks
(artefact\footnotemark[1]) confirm reproducible convergence (cross-seed
loss std $<1\%$), identical LoRA parameter counts per model, and the
expected canary-visit budget of 30 (deterministic in V0--V1, Poisson with
$\sigma\!\approx\!5.5$ in V2--V3, averaged over three seeds).

Table~\ref{tab:exp1_main} summarizes cross-seed results. V0$\to$V1 reduces
mean exposure by $36$--$66\%$ (Gemma~$-60.5\%$, Qwen3~$-36.8\%$,
Llama~$-35.7\%$, VaultGemma~$-65.6\%$), with Cliff's
$\delta\in[-0.60,-0.38]$ and $p<10^{-5}$ in all Wilcoxon tests. Neither
V1$\to$V2 ($+\varepsilon{=}8$) nor V2$\to$V3
($\varepsilon{=}8\to2$) is significant after FDR correction.

\noindent\textit{Class asymmetry.}
Splitting 20 canaries into Class~A (anomalous) and Class~B (realistic),
V0 shows higher exposure for Class~B in three models (Gemma $3.70$ vs.\
$3.09$; Qwen3 $4.88$ vs.\ $3.85$; VaultGemma $3.88$ vs.\ $3.32$; Llama
balanced). This contradicts the expectation that documentation-range
identifiers leak more: IPs such as \texttt{203.0.113.42} are familiar
from pretraining and receive similar likelihood to nearby
\texttt{/24} variants, whereas realistic identifiers (e.g.,
\texttt{prod-srv-042.interno.rnp.br}) involve rarer token combinations,
consistent with duplication-frequency--memorization
scaling~\cite{Kandpal2022}. The asymmetry attenuates in V1--V3,
indicating roughly uniform effects of the stack and DP-SGD (V0 ordering:
asset $>$ FQDN $>$ IPv6 $>$ IPv4 $>$ MAC). Realistic identifiers are
thus more vulnerable targets.

\begin{table}[!htp]
\centering
\caption{Experiment~1 main results, pooled across 3 seeds. Lower exposure and AUC closer to $0.5$ are better.}
\label{tab:exp1_main}
\begin{tabular}{lcccc}
\toprule
\textbf{Model} & \textbf{V0} & \textbf{V1} & \textbf{V2 ($\varepsilon{=}8$)} & \textbf{V3 ($\varepsilon{=}2$)} \\
\midrule
\multicolumn{5}{l}{\emph{Mean exposure (lower is better)}} \\
Gemma 3 1B      & 3.49 & 1.38 & 1.54 & 1.50 \\
Qwen3 1.7B      & 4.48 & 2.83 & 2.60 & 2.51 \\
Llama 3.2 3B    & 4.98 & 3.20 & 2.68 & 2.92 \\
VaultGemma 1B   & 3.60 & 1.24 & 1.26 & 1.45 \\
\midrule
\multicolumn{5}{l}{\emph{AUC of loss-canary (0.5 = random)}} \\
Gemma 3 1B      & 0.592 & 0.510 & 0.522 & 0.525 \\
Qwen3 1.7B      & 0.770 & 0.570 & 0.564 & 0.558 \\
Llama 3.2 3B    & 0.848 & 0.594 & 0.578 & 0.576 \\
VaultGemma 1B   & 0.596 & 0.486 & 0.487 & 0.494 \\
\bottomrule
\end{tabular}
\end{table}

\noindent\textit{Decomposition: optimizer updates vs.\ DP-SGD.}
V0 (lot~$4$, $\sim$$2400$ updates over 3 epochs) and V1 (lot~$32$, $\sim$$300$
updates over 3 epochs) differ in three coupled axes: lot size, NF4
quantization, and total optimizer updates. Matched-update controls
(V0-FewStep: V0 truncated to $\sim$$300$ updates with all other settings
held fixed) attribute the dominant share to update count: cross-seed
($n{=}3$) collapses reproduce $66$--$132\%$ of the V0$\!\to\!$V1 gap with
only fewer updates as the changing variable (mean $100\%$ across 4 models;
Gemma $89\%$, Qwen $114\%$, Llama $132\%$, VaultGemma $66\%$), in
line with update-driven memorization scaling~\cite{Carlini2019,Hoffer2017}.
Head-to-head comparisons of raw fine-tuning versus
DP-SGD~\cite{Yu2022,Li2022,Lukas2023} therefore attribute to DP-SGD a
reduction that update count predicts: in our setting the V0$\!\to\!$V1
effect (mean $\Delta{\approx}{-}1.98$ bits, range $-1.65$ to $-2.37$) is
largely a consequence of $8\times$ fewer optimizer updates at lot~$32$,
while DP-SGD contributes the formal $(\varepsilon,\delta)$ guarantee plus
a marginal further reduction in the models with largest residual V1
exposure.

\noindent\textit{Stack ablation: lot size vs.\ NF4 quantization.}
Within the V0$\to$V1 stack we isolate the marginal effects of large-lot batching and 4-bit quantization using two intermediates: $A$ (lot~32, bf16, uniform shuffle) and $B$ (lot~4, NF4, uniform shuffle), trained under the same protocol as V0/V1 across all four models, cross-seed (n${=}3$). A third cell, $C$ (lot~4, Poisson, bf16), was infeasible on the 12--16~GB GPUs of \S\ref{sec:models} without Opacus \texttt{BatchMemoryManager} (out-of-memory at cross-entropy and empty-batch failures), preventing separation of lot size and sampling effects. The joint $\Delta_{V1}$ is robustly negative across all four models ($\in[-2.37,-1.65]$~bits, all $95\%$ paired bootstrap CIs over $60$ canary$\times$seed observations exclude zero, $B{=}10{,}000$); the lot-vs-NF4 split is architecture-dependent. On Gemma~3~1B and Llama~3.2~3B, lot size alone reproduces $\sim$$95$--$100\%$ of $\Delta_{V1}$ ($\Delta\_A{=}{-}2.00,{-}1.79$) with $\Delta\_B$ small ($\Delta\_B{=}{-}0.67,{-}0.19$). On Qwen3~1.7B, lot accounts for $\sim$$93\%$ ($\Delta\_A{=}{-}1.53$, $\Delta\_B{=}{-}0.63$). On VaultGemma~1B the ordering inverts: NF4 dominates ($\Delta\_B{=}{-}1.41$, $\Delta\_A{=}{-}0.96$). Within the residual stack effect (i.e., once step-count attribution is removed), the lot-vs-NF4 split is model-conditional. LoRA-related reductions are also observed in FL~\cite{Bossy2025LoRAFL}.
The matched-update result above further generalizes the V0$\!\to\!$V1 attribution cross-seed and resolves the $8{\times}$-step confound flagged earlier (\S\ref{sec:limitations}).

\section{Pseudonymization as Preprocessing}
\label{sec:exp2}

The second experiment mirrors the first one with one change: the dataset is
processed by \anontool{} (Section~\ref{sec:datasets}) before training. The
variants Anon-V0--Anon-V3 are paired with V$k$ and evaluated under two
extraction attacks: \emph{Attack~1} queries the original canary (does
substitution prevent memorization?), and \emph{Attack~2} queries the slug
(does memorization shift to the substitute?). Audit details are in the
artefact.

\noindent\textit{Attack 1: original-canary exposure under pseudonymization.}
Table~\ref{tab:exp2_main} compares original-canary exposure between V0
(Experiment~1) and Anon-V0, isolating pseudonymization without the
protection stack. HMAC reduces exposure by $40$--$61\%$ across models,
with residual means $1.31$--$2.60$ bits (floor $\sim$1~bit), indicating
attenuation but not elimination. For V1--V3, the marginal reduction is
within $\pm 5\%$, as baselines already sit at the floor. 

\begin{table}[!htp]
\centering
\caption{Effect of pseudonymization on exposure of the original canary, V0 regime. Values are \emph{per-class means} (Section~\ref{sec:canaries}) over 3 seeds; numbers therefore differ slightly from the per-canary means in Tab.~\ref{tab:exp1_main}. Reduction is $\mathrm{exp}(V0) - \mathrm{exp}(\text{Anon-V0})$, normalized by $\mathrm{exp}(V0)$.}
\label{tab:exp2_main}
\begin{tabular}{lcccc}
\toprule
\textbf{Model} & \textbf{exp(V0)} & \textbf{exp(Anon-V0)} & \textbf{Abs. reduction} & \textbf{Rel. reduction} \\
\midrule
Gemma 3 1B      & 3.394 & 1.311 & $+2.083$ & $+61.4\%$ \\
Qwen3 1.7B      & 4.362 & 2.597 & $+1.764$ & $+40.5\%$ \\
Llama 3.2 3B    & 4.866 & 2.157 & $+2.708$ & $+55.7\%$ \\
VaultGemma 1B   & 3.601 & 1.397 & $+2.204$ & $+61.2\%$ \\
\bottomrule
\end{tabular}
\end{table}

\noindent\textit{Attack 2: slug memorization.}
\label{sec:exp2_slug}
The dual protocol completes by querying for the HMAC slug itself (e.g.,
\texttt{[IP\_ADDRESS\_6975bab4]}, $10$--$16$ tokens depending on
tokenizer). Variants in the pool are HMAC-substituted with the same key,
so each entry has the same surface form
$\mathtt{[}\textit{TYPE}\mathtt{\_}\langle\textit{8 hex}\rangle\mathtt{]}$
as the planted slug. Table~\ref{tab:exp2_attack2} reports cross-seed mean
slug exposure across all 16 (model$\times$variant) cells. The result is
sharp: slug exposure stays in $[0.88, 1.63]$ bits, within $\pm 0.65$ bits
of the floor. Qwen3 1.7B and Llama 3.2 3B, which produced the highest
original-canary exposure in Experiment~1 ($4.48$ and $4.98$ bits at V0),
give slug exposures of $0.88$--$1.37$ bits, substantially \emph{lower}
than their original-canary exposure on the Anon-V$k$ adapters
($1.94$--$2.61$ bits). Gemma and VaultGemma sit near floor in all
variants.

\begin{table}[!htp]
\centering
\caption{Mean exposure (bits) of the HMAC slug under Attack~2, cross-seed (mean $\pm$ std over 3 seeds). Random baseline is $\log_2(101/50) \approx 1.0$ bit. All 16 cells stay within $\pm 0.65$ bits of the floor.}
\label{tab:exp2_attack2}
\begin{tabular}{lcccc}
\toprule
\textbf{Model} & \textbf{Anon-V0} & \textbf{Anon-V1} & \textbf{Anon-V2 ($\varepsilon{=}8$)} & \textbf{Anon-V3 ($\varepsilon{=}2$)} \\
\midrule
Gemma 3 1B    & $1.497 \pm 0.248$ & $1.494 \pm 0.212$ & $1.337 \pm 0.418$ & $1.440 \pm 0.275$ \\
Qwen3 1.7B    & $1.276 \pm 0.664$ & $1.026 \pm 0.319$ & $0.880 \pm 0.174$ & $0.918 \pm 0.174$ \\
Llama 3.2 3B  & $0.993 \pm 0.351$ & $1.014 \pm 0.310$ & $1.373 \pm 0.266$ & $1.337 \pm 0.161$ \\
VaultGemma 1B & $1.552 \pm 0.412$ & $1.634 \pm 0.240$ & $1.360 \pm 0.176$ & $1.506 \pm 0.133$ \\
\bottomrule
\end{tabular}
\end{table}

HMAC-SHA256 truncated to 8 hex chars is computationally indistinguishable
from uniform random without the key (HMAC's PRF
property~\cite{Krawczyk1997RFC2104,Bellare2006}); cryptographic
unpredictability ensures the variant pool is a~priori indistinguishable
from the planted slug, but does not by itself preclude memorization of the
specific observed value during training. The empirical contribution is
that, across all 16 cells, no extractable preference for the planted slug
emerges. Constructing the same-keyed variant pool requires the HMAC key
and is therefore an evaluation tool: a worst-case audit, not a black-box
attack a real adversary could mount, since under our threat model
(Section~\ref{sec:threat}) the key is not exposed. Combined with
Attack~1's $40$--$61\%$ reduction at V0 (Table~\ref{tab:exp2_main}), HMAC
pseudonymization removes the original from the training-time exposure
surface without creating an extractable secondary memorization target.

\section{Utility Evaluation}
\label{sec:utility}

\subsection{Logit-Based Classification on Fine-Tuned Adapters}
\label{sec:utility_adapter}

We evaluate downstream utility through severity classification on 600
held-out vulnerability records using CVSS buckets (Low, Medium, High,
Critical) per the FIRST CVSS~v3.1 specification~\cite{CVSS31}. Following
HELM-style multiple-choice evaluation~\cite{Liang2022HELM}, we use
logit-based scoring rather than free-form generation. For a record $x$ and
candidate label $\ell\in\{\mathrm{Low,Med,High,Crit}\}$,
\begin{equation}
S(\ell\mid x)=\sum_{i=1}^{|c_{\ell}|}\log P_{\theta}\bigl(c_{\ell,i}\mid
\mathrm{prompt}(x), c_{\ell,1:i-1}\bigr),
\label{eq:logit_score}
\end{equation}
with $c_{\ell}$ the canonical token continuation of label $\ell$ and
$\hat\ell(x)=\arg\max_{\ell}S(\ell\mid x)$.

\noindent\textit{Prompt-format sensitivity.}
Among four prompt formats evaluated on V0 adapters
(Table~\ref{tab:utility_ablation}: training-faithful, truncated pre-CVSS,
closed-dict, natural language), training-faithful is the best or
near-best format on all four models, with cross-format variation up to
$2.75\times$ in F1-macro, consistent
with prompt-format sensitivity in instruction-tuned LMs~\cite{Sclar2024};
in our setting, the JSON-faithful training distribution does not transfer
to instruction-style prompt formats. Logit-based scoring with
training-faithful prompts is therefore mandatory for evaluating
autoregressive SLMs fine-tuned over structured data; an initial
generation-based pilot produced $\sim$95\% invalid outputs and was
abandoned.

\begin{table}[!htp]
\centering
\caption{\textbf{[Fine-tuned models]} Format ablation on V0 LoRA adapters (ablation pilot, single seed, 50 held-out records per cell). F1-macro reported.}
\label{tab:utility_ablation}
\resizebox{.90\linewidth}{!}{%
\begin{tabular}{lcccc}
\toprule
\textbf{Format} & \textbf{Gemma 3 1B} & \textbf{Qwen3 1.7B} & \textbf{Llama 3.2 3B} & \textbf{VaultGemma 1B} \\
\midrule
Training-faithful  & 0.247 & 0.240 & \textbf{0.324} & 0.124 \\
Truncated pre-CVSS & 0.087 & 0.191 & 0.252 & 0.114 \\
Closed-dict        & 0.194 & 0.259 & 0.241 & 0.148 \\
Natural language   & 0.209 & 0.145 & 0.152 & 0.169 \\
\bottomrule
\end{tabular}
}
\end{table}

\noindent\textit{Cross-seed results.}
Table~\ref{tab:utility_adapter} reports the full 96-adapter scan with three
metrics (F1-macro, top-1 accuracy, adjacent accuracy at
$|y-\hat y|\le 1$). Two baselines bound interpretation: uniform random
(F1$\approx$0.20, Acc$=$0.250, Adj$=$0.625) and majority-class always-Medium
(F1$=$0.125, Acc$=$0.333, Adj$=$0.833). F1-macro stays in $[0.19, 0.28]$
across all 96 cells, marginally above the always-Medium F1 baseline of
$0.125$ but far below an operationally useful threshold ($\geq 0.5$).
Top-1 accuracy peaks at Llama V0 ($0.30$), still under the always-Medium
$0.333$. Adjacent accuracy peaks at Llama V0 ($0.716$, $\sim$0.09 above
the random baseline), with no cell reaching the always-Medium ordinal
$0.833$. Within each model, V0/V1/V2/V3 differ by $\leq 0.07$ on F1-macro;
a per-record paired Wilcoxon of V$k$ vs.\ Anon-V$k$ yields no significant
separation after BH-FDR. Severity classification serves here as a
\emph{probe}, not as an operationally needed task (CSIRT pipelines already
derive CVSS severity from CVE metadata).

\begin{table}[H]
\centering
\caption{\textbf{[Fine-tuned models]} Cross-seed LoRA adapter utility on the 600-record held-out, training-faithful format. F1 = F1-macro; Acc = top-1 accuracy; Adj = adjacent accuracy ($|y - \hat{y}| \leq 1$ on the bucket index). All values are means over 3 seeds. Random uniform baselines: F1$\approx$0.20, Acc=0.250, Adj=0.625. Majority-class (always Medium) baselines: F1=0.125, Acc=0.333, Adj=0.833.}
\label{tab:utility_adapter}
\setlength{\tabcolsep}{4pt}
\scriptsize
\begin{tabular}{llcccccccc}
\toprule
 & & \multicolumn{2}{c}{V0} & \multicolumn{2}{c}{V1} & \multicolumn{2}{c}{V2 ($\varepsilon{=}8$)} & \multicolumn{2}{c}{V3 ($\varepsilon{=}2$)} \\
\cmidrule(lr){3-4} \cmidrule(lr){5-6} \cmidrule(lr){7-8} \cmidrule(lr){9-10}
\textbf{Model} & \textbf{Metric} & raw & anon & raw & anon & raw & anon & raw & anon \\
\midrule
\multirow{3}{*}{Gemma 3 1B}
 & F1  & 0.250 & 0.251 & 0.224 & 0.227 & 0.228 & 0.242 & 0.232 & 0.241 \\
 & Acc & 0.273 & 0.267 & 0.251 & 0.234 & 0.257 & 0.247 & 0.260 & 0.246 \\
 & Adj & 0.657 & 0.643 & 0.641 & 0.577 & 0.647 & 0.592 & 0.646 & 0.584 \\
\midrule
\multirow{3}{*}{Qwen3 1.7B}
 & F1  & 0.257 & 0.246 & 0.213 & 0.212 & 0.206 & 0.219 & 0.202 & 0.217 \\
 & Acc & 0.279 & 0.269 & 0.264 & 0.254 & 0.249 & 0.257 & 0.246 & 0.254 \\
 & Adj & 0.673 & 0.664 & 0.684 & 0.676 & 0.683 & 0.681 & 0.683 & 0.682 \\
\midrule
\multirow{3}{*}{Llama 3.2 3B}
 & F1  & \textbf{0.278} & 0.260 & 0.210 & 0.246 & 0.246 & 0.231 & 0.245 & 0.241 \\
 & Acc & \textbf{0.303} & 0.281 & 0.241 & 0.273 & 0.258 & 0.249 & 0.254 & 0.254 \\
 & Adj & \textbf{0.716} & 0.698 & 0.671 & 0.693 & 0.624 & 0.681 & 0.614 & 0.679 \\
\midrule
\multirow{3}{*}{VaultGemma 1B}
 & F1  & 0.212 & 0.234 & 0.189 & 0.253 & 0.196 & 0.250 & 0.198 & 0.258 \\
 & Acc & 0.249 & 0.259 & 0.224 & 0.262 & 0.242 & 0.267 & 0.242 & 0.274 \\
 & Adj & 0.595 & 0.643 & 0.601 & 0.661 & 0.601 & 0.684 & 0.607 & 0.693 \\
\bottomrule
\end{tabular}
\end{table}

\noindent\textit{Viability gap under fixed budget.}
Adjacent CSIRT classification literature on anonymized
data~\cite{AnonRef1} sharpens the reading: \cite{AnonRef1} reports
$96.4\%$ top-bucket match (cosine $>$0.8) on a 12-class
NIST~SP~800-61r3-derived~\cite{NIST80061r3} benchmark with Gemini~2.0~Flash, and~\cite{AnonRef2} reports
$\sim$53\% (7--12B) and $\sim$61.7\% (20--70B) accuracy with prompt
engineering on local SLMs. Combined with our prompt-engineering result on
the same base models (\S\ref{sec:utility_prompt},
Table~\ref{tab:utility_prompt}; F1$\in[0.00,0.17]$), the bilateral
evidence is consistent with a budget-conditional gap: under the evaluated
training regime ($3$ epochs over $3{,}200$ records, no convergence-based
stopping), $1$--$3$B SLMs do not reach operational F1 on this task, while
$\geq$7B models with prompt engineering on adjacent narrative data sit
substantially higher (different tasks). Whether the gap closes with
longer training, larger models, or different task formulations is left
to future work (\S\ref{sec:limitations}).

\subsection{Prompt Engineering on Base Models}
\label{sec:utility_prompt}

To complement the adapter result, we evaluate raw vs.\ Anon data on base
instruction-tuned models with a four-shot severity prompt~\cite{Brown2020GPT3}
(one midrange example per CVSS bucket) over the 600 held-out records,
following~\cite{AnonRef1}. Pseudonymization preserves utility on
the three instruction-tuned models (Table~\ref{tab:utility_prompt};
$|\Delta F_1|<0.05$ in every case; BH-FDR-significant for Gemma and Qwen
with non-negative direction). VaultGemma is the documented exception: as a
DP-pretrained release without instruction tuning it produces $100\%$
invalid outputs in both versions, so $\Delta F_1{=}0$ is non-functioning
rather than utility preservation. Replacing structured identifiers by
HMAC slugs does not destroy the information needed for classification,
which depends on free-text descriptions, CVE/CWE references, and
statistical patterns rather than on the substituted identifiers.

\begin{table}[!htp]
\centering
\caption{\textbf{[Base models, NO fine-tuning]} Severity classification on 600 held-out records, four-shot prompt, base instruction-tuned models. $F_1$ is macro-averaged over four CVSS buckets. Wilcoxon is paired by record (anon\_correct $-$ raw\_correct $\in \{-1, 0, +1\}$); BH-FDR over four comparisons.}
\label{tab:utility_prompt}
\resizebox{.85\textwidth}{!}{%
\begin{tabular}{lcccccc}
\toprule
\textbf{Model} & $F_1$\textbf{ raw} & $F_1$\textbf{ anon} & $\Delta F_1$ & \textbf{Inv. raw} & \textbf{Inv. anon} & \textbf{BH-FDR} \\
\midrule
Gemma 3 1B    & 0.136 & 0.146 & $+0.010$ & 0.0\%  & 0.0\%  & $\checkmark$ \\
Qwen3 1.7B    & 0.165 & 0.167 & $+0.002$ & 29.7\% & 11.0\% & $\checkmark$ \\
Llama 3.2 3B  & 0.071 & 0.077 & $+0.006$ & 0.0\%  & 0.0\%  & --           \\
VaultGemma 1B & 0.000 & 0.000 & $0.000$  & 100\%  & 100\%  & --           \\
\bottomrule
\end{tabular}
}
\end{table}

\section{Discussion}
\label{sec:discussion}

The two defenses address orthogonal threats and live at different points
in the pipeline. Pseudonymization removes the original from the training
distribution (the model never sees it) and operates as a data-side
preprocessing pass with no model coupling. DP-SGD bounds per-example
influence on the final weights and is coupled to the training loop, the
autograd graph, and (in PyTorch) the Opacus per-sample-gradient hooks.
DP-SGD therefore retains independent value through the formal
$(\varepsilon,\delta)$ guarantee that compliance frameworks may require,
even when its empirical contribution to memorization reduction is
dominated by reduced optimizer-update count (\S\ref{sec:exp1}). VaultGemma's
DP-at-pretraining~\cite{Sinha2025} adds a third axis but, consistent with
Tram\`er et al.~\cite{Tramer2022}, bounds leakage of pretraining data
only; fine-tuning-time canaries lie outside its scope, as our V0/V1
exposures on VaultGemma confirm.

\noindent\textit{Deployment.} As the open-weight SLM ecosystem moves
toward multimodal-by-default architectures, DP-SGD on Gemma~4-class or
Qwen3-VL-class models requires architecture-specific LoRA
module-exclusion regexes (\S\ref{sec:models}), whereas a pseudonymization
layer adds no architecture-dependent cost. CSIRTs targeting on-premise
classification may benefit from looking above the 3B regime where our
budget-conditional gap was observed; pseudonymization remains useful regardless of
model size and composes with whichever fine-tuning-time defense is selected.

\section{Limitations}
\label{sec:limitations}

\emph{Dataset:} Mock~CAIS diverges from Tenable in long-form text
(\S\ref{sec:datasets}) and lacks multi-scan lineage.
\emph{Models:} four text-only $1$--$3$B SLMs under 16~GB VRAM; multimodal
SLMs are excluded by the DP-SGD constraint of \S\ref{sec:models}, and
$\geq$7B is future work. \emph{Probe task:} severity classification is
one probe; the budget-conditional gap is supported bilaterally
(\S\ref{sec:utility}) and by the $\geq$7B envelope of
\cite{AnonRef2}, but a task sweep remains open.
\emph{Threat model:} black-box log-probability only; white-box gradient
leakage~\cite{Geiping2020}, 
model extraction~\cite{Carlini2024Stealing},
training-time MIA~\cite{Shokri2017}, and distribution-shift confounds in
loss-based MIA~\cite{Duan2024MIA} are out of scope.
\emph{Coverage:} \anontool{} is schema-driven, so residual identifiers in
\texttt{description}/\texttt{solution} are not measured; lot size, LoRA
rank, and $K{=}10$ replicas are fixed.
\emph{Ablation:} the lot-vs-NF4 stack split (\S\ref{sec:exp1}) is reported
cross-seed (n${=}3$); cell $C$ (lot~4, Poisson, bf16) hit OOM on $12$--$16$~GB
GPUs without the \texttt{BatchMemoryManager} layer of the DP-SGD path,
preventing separation of lot size from sampling. We report matched-update
controls cross-seed (\S\ref{sec:exp1}); equal-step controls under linear
LR scaling across lot sizes are deferred~\cite{Marek2025SmallBatch}. The
Opacus + LoRA + quantization incompatibility persists in versions 1.5 and
1.6 for lot $<$ 32 and is reported as a deployment hazard.
\emph{Privacy auditing:} Opacus PRV
$\varepsilon$ only; single-run auditing
and probabilistic extraction
are future work.

\section{Conclusion}
\label{sec:conclusion}

Across 96 LoRA adapters and a dual extraction attack on HMAC
pseudonymization, three findings emerge. First, matched-update controls
attribute $66$--$132\%$ of the V0$\!\to\!$V1 memorization reduction (mean $100\%$)
($\Delta\!\approx\!{-}1.98$ bits) to fewer optimizer updates rather than
to the stack itself, while DP-SGD adds the formal $(\varepsilon,\delta)$
guarantee without significant further reduction. Second, cryptographic
pseudonymization is architecture-agnostic and removes the original
identifier from the exposure surface, with HMAC slug exposure within
$\pm0.65$ bits of the metric floor across all 16 (model$\times$variant)
cells, i.e., no secondary memorization target emerges. Third, under the
evaluated training budget, $1$--$3$B SLMs do not reach operational F1.
Future work: factorial decomposition (lot$\times$quant$\times$DP) under
matched updates on an expanded corpus, a dual keyed/keyless slug threat
model, and toolchain fixes for the persistent Opacus + LoRA + quantization
incompatibility (lot $<$ 32, versions 1.5--1.6).
%
%
\bibliographystyle{plain}
\bibliography{references}

\end{document}